\documentclass[twocolumn,prl,showpacs,preprintnumbers,amsmath,amssymb]{revtex4-1}
\usepackage{graphicx}
\usepackage{dcolumn}
\usepackage{bm}
\usepackage{color}
\usepackage[applemac]{inputenc}
\usepackage{bm}
\usepackage{amsmath}
\usepackage{amsfonts}
\usepackage{amssymb}
\usepackage{amsthm}
\usepackage{amstext}
\usepackage{amsbsy}
\usepackage{amsopn}
\usepackage{amscd}
\usepackage{amsxtra}

\def\phi{\varphi}
\def\epsilon{\varepsilon}

\everymath{\displaystyle}

\renewcommand{\Re}{\operatorname{Re}}

\begin{document}
\title{Orientation and Alignment  Echoes }

\author{G. Karras$^1$}
\author{E. Hertz$^1$}
\author{F. Billard$^1$}
\author{B. Lavorel$^1$}
\author{J.-M. Hartmann$^2$}
\author{O. Faucher$^1$}
\email{olivier.faucher@u-bourgogne.fr}
\affiliation{$^1$Laboratoire Interdisciplinaire CARNOT de Bourgogne, UMR 6303 CNRS-Universit\'e de Bourgogne, BP 47870, 21078 Dijon, France}
\affiliation{$^2$Laboratoire Interuniversitaire des Syst\`{e}mes Atmosph\'{e}riques (LISA) CNRS (UMR 7583), Universit\'{e} Paris Est Cr\'{e}teil, Universit\'{e} Paris Diderot, Institut Pierre-Simon Laplace, Universit\'{e} Paris Est Cr\'{e}teil, 94010 Cr\'{e}teil Cedex, France}
\author{Erez Gershnabel}
\affiliation{Department of Chemical Physics, The Weizmann Institute of Science, Rehovot 76100, Israel}
\author{Yehiam Prior}
\affiliation{Department of Chemical Physics, The Weizmann Institute of Science, Rehovot 76100, Israel}
\author{Ilya Sh. Averbukh}
\email{ilya.averbukh@weizmann.ac.il}
\affiliation{Department of Chemical Physics, The Weizmann Institute of Science, Rehovot 76100, Israel}
\newlength{\textlarg}
\newcommand{\strike}[1]{%
  \settowidth{\textlarg}{#1}
  #1\hspace{-\textlarg}\rule[0.5ex]{\textlarg}{0.5pt}}

\date{\today}
\begin{abstract}
We present  what is probably the simplest classical system featuring the echo phenomenon - a collection of randomly oriented free rotors  with dispersed rotational velocities. Following excitation  by a pair of time-delayed impulsive kicks, the mean orientation/alignment of the ensemble exhibits multiple echoes and fractional echoes. We elucidate the mechanism of the echo formation by   kick-induced  filamentation of phase space, and provide the  first experimental demonstration of classical alignment echoes in a thermal gas of CO$_2$ molecules excited by a pair of femtosecond laser pulses.
\end{abstract}

\maketitle
Echoes are common in many areas of physics. When an inhomogeneous ensemble of many nonlinear systems is impulsively kicked by  an external force, the impulsive response of the ensemble typically disappears soon after the end of the stimulating pulse. While each individual sub-system maintains its coherence after the kick, the total response, as quantified by some ensemble-averaged quantity, vanishes due to the dispersion in the properties of the individual systems. If the  ensemble is stimulated  again by a second, delayed kick, a transient response appears and disappears fast, but a new impulsive response (echo) shows up at twice the delay between the two pulses. The most celebrated examples of this phenomenon are the Hahn and Carr-Purcell echoes of precessing nuclear spins \cite{Hahn,Carr}. Echoes have been observed in a wide class of classical systems, including  cyclotron echo \cite{Hill,Gould},  plasma wave echo \cite{plasma}, and photon echo \cite{photon}. Echoes were predicted to occur in  proton storage rings~\cite{Stupakov-h1, Stupakov-h2} and they were observed in high energy hadron beam  experiments at Fermilab \cite{Fermilab} and CERN \cite{CERN}. More recently, echo-enabled generation of short-wavelength radiation in free-electron lasers \cite{Stupakov,Xiang,Zhao} was demonstrated. On the opposite side of the  energy spectrum, echoes are being discussed in the context of cavity quantum electrodynamics \cite{Walther,Haroche} and cold atom systems~ \cite{Rabitz,Ertmer,Davidson,Fishman}.

Despite the general character of the echo phenomenon, an all-encompassing physical qualitative description of the echo mechanism does not exist. In the case of the nuclear spin echo \cite{Hahn,Carr}, the effect is intuitively  explained by  time reversal of the quantum spin dynamics on the Bloch sphere. However, in most other appearances  of echoes, no such transparent explanation exists, and the analysis typically relies on mathematical arguments to reveal nonlinear phase inversion in the response signal.

Here we identify what is probably the simplest classical system featuring the echo phenomenon - a collection of identical free classical rotors stimulated by an external impulsive force.  Using only geometric arguments relating to the phase-space transformations of free  rotation  subject  to impulsive  kicks,  we predict multiple echoes in the mean orientation/alignment of the ensemble. Our qualitative analysis not only reveals the mechanism of these echoes in terms of a very simple model, but also predicts the behavior of the echo signal as a function of the stimulating kicks parameters.

A full analytical theory of these echoes in 2D and 3D thermal ensembles of classical rotors, and its generalization to the quantum case will be published elsewhere.  Here, we discuss the mechanism behind the predicted echo effect and demonstrate it experimentally in a collection of CO$_2$ molecules stimulated by a pair of femtosecond laser pulses.

 For a linear molecule having a permanent dipole moment $\mu$, and driven by a linearly polarized field, the interaction potential leading to orientation is
\begin{equation}
V(\theta ,t)=-\mu E(t) \cos (\theta ),
\label{dipole}
\end{equation}
where $E(t)$  is the field amplitude, and  $\theta$ is the polar angle between the molecular axis and the field direction.  In the absence of  a permanent dipole moment, the external field couples to the induced molecular polarization. For nonresonant laser fields, this interaction, averaged over fast optical oscillations, is \cite{Boyd,Friedrich}
\begin{equation}
V(\theta ,t)=-\frac{1}{4} E^2 (t) [(\alpha_{\|}-\alpha_{\bot})\cos^2 (\theta )+\alpha_{\bot}],
\label{polarizable}
\end{equation}
which leads to alignment along the field polarization (for reviews on laser molecular alignment see \cite{Tamar,Ohshima10,Fleischer12,Krems,Pabst13}). Here $\alpha_{\|}$  and $\alpha_{\bot}$  are the components of the polarizability, parallel and perpendicular to the molecular axis, and $E(t)$ is the {\it envelope} of the laser pulse. Although these two  forms of $V(\theta ,t)$  lead to different physical consequences (i.e., orientation vs alignment), the effects we   present  here are rather insensitive to the choice of interaction.
Thus, in what follows, we will use for discussion the dipole interaction form of Eq.(\ref{dipole}), responsible for orientation, and will point out several differences appearing in the case of alignment.

\begin{figure}
\centering \includegraphics[width=0.4\textwidth]{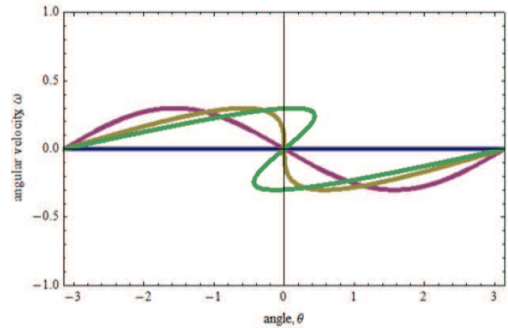}
\protect\caption{(color online) Kick-induced orientation of an ensemble of rotors.}
\label{orientation}
\end{figure}

We start with a  description of the impulsive field-free orientation of an ensemble of 2D  rotors kicked by a single short pulse at $t=0$. The angular velocity, $\omega$ and angle $\theta$ of a rotor at time $t$  are given by
\begin{eqnarray}\label{map}
\omega &=& \omega_0-\Omega \sin (\theta_0 ) \\
\nonumber \theta &=& \theta_0 +\omega t
\end{eqnarray}
Here $\omega_0, \theta_0$ are the initial conditions, and $\Omega$ is proportional to the intensity of the kick. The area-preserving transformation (\ref{map}) is  (up to change in  notations) the so called Chirikov-Taylor map that is widely used in the studies on deterministic chaos \cite{chaos}.
The  orientation of the ensemble of rotors is quantified by the mean value of $\langle\cos(\theta ) \rangle$, referred to as the orientation factor.
For simplicity, let us first assume that all the rotors are initially at rest and uniformly distributed in the angular interval $[-\pi,\pi]$, as represented by the horizontal blue line in Fig.\ref{orientation}.
A short kick of force (delta kick) in the $\theta=0$  direction does not move the rotors during its action, but induces rotation with  angular speed  $-\Omega \sin (\theta )$  (see Eq.(\ref{map})).  Immediately after the kick, the distribution of the rotors in  phase space takes a shape shown by the red curve in Fig.\ref{orientation}.
 In the course of the following free evolution, each point on  the red curve moves horizontally
towards  $\theta=0$ with a velocity  defined  by its  initial vertical position, as shown  by the next two curves.
After a certain delay, the curve experiences steepening at $\theta=0$  (light green curve), not unlike the accumulation of cars in congested traffic, which leads to a singularity in the angular distribution of the rotors \cite{AA}. Note, however, that the maximal value of the orientation factor $\langle\cos(\theta ) \rangle$  is achieved not at this moment, but some time later, when the curve takes a typical folded shape (see dark-green curve in Fig.\ref{orientation}) leading to  increased density of  rotors in the  region near $\theta=0$  \cite{AA}. This is a transient orientation, and the stronger the kick the shorter is the time needed to reach the maximally oriented state.

We now switch gears, and consider an ensemble of kicked rotors that is initially uniformly dispersed  in the  interval  $[-\pi,\pi]$, but  with  a spread  in angular velocity $\omega$.  We assume a Gaussian distribution  of angular velocities $f(\theta_0, \omega_0 )\sim \exp[-\omega_0^2/(2\sigma^2)]$. Here $\sigma$ may represent, for instance,  a  standard deviation of the angular velocity in a thermal ensemble. The time-dependent orientation factor after the kick is found to be: $\langle \cos (\theta ) (t)\rangle = \exp \left[-\sigma^2 t^2/2 \right] J_1 (\Omega t)$, where $J_1(z)$ is the Bessel function of the first order. This pulsed response disappears fast as $\sigma t\rightarrow \infty$. Deeper insight can be obtained from the analysis of the time-dependent distribution function in  phase space. By inverting the map (\ref{map}) and using its area preserving property, one arrives at the following expression for the probability distribution function at time $t$:
\begin{equation}
  f(\omega ,\theta ,t)=\frac{1}{2\pi}\frac{1}{\sqrt{2\pi}\sigma}\exp\left[-\frac{[\omega-\Omega\sin (\omega t-\theta)]^2}{2\sigma^2}\right]
  \label{distfun}
\end{equation}

\begin{figure}
\centering \includegraphics[width=0.5\textwidth]{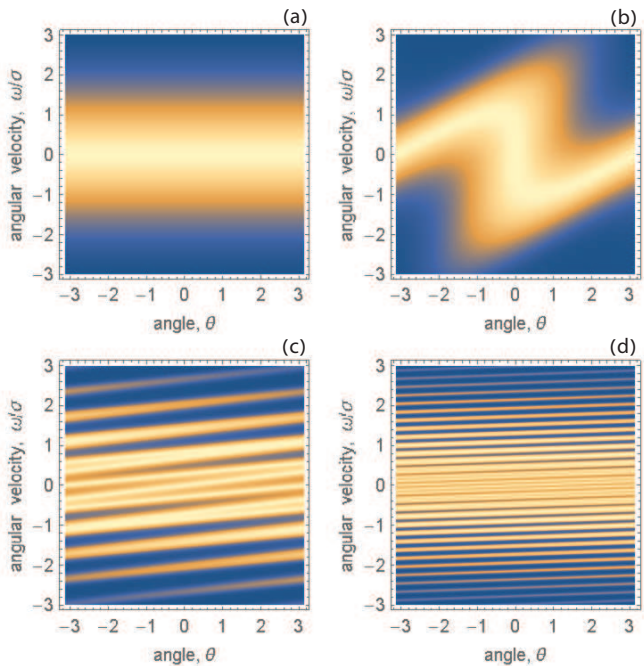}
\protect\caption{(color online) Filamentation of the phase space density distribution. $\Omega/\sigma=1$, (a) initial distribution at $t=0$, (b) $\sigma t=1$, (c) $\sigma t=10$, (d) $\sigma t=30$.}
\label{firstkick}
\end{figure}

Figures \ref{firstkick}a-d show the gradual transformation of the initial distribution (Fig. \ref{firstkick}a) with time. Shortly after the kick (Fig. \ref{firstkick}b) the density distribution takes a folded shape similar to the one shown in Fig.~\ref{orientation}.  On the longer time scales, when the orientation signal $\langle \cos (\theta ) \rangle (t)$ vanishes,  the probability density becomes rippled and develops  multiple parallel filaments  (see Figs.~\ref{firstkick}c,d). The number of these filaments grows with time, and their width is diminishing in order to keep the occupied phase space volume constant.  Eventually, all the filaments tend to become almost horizontal and uniform in density.   The filamentation of the phase space is a phenomenon well known in accelerator physics \cite{Lichtenberg}, and in statistical mechanics of stellar systems \cite{stellar}. In our case (as follows from Eq.(\ref{distfun})), the neighboring filament strips are separated in angular velocity by  $2\pi/t$, where $t$  is the evolution time.

\begin{figure}
\centering \includegraphics[width=0.5\textwidth]{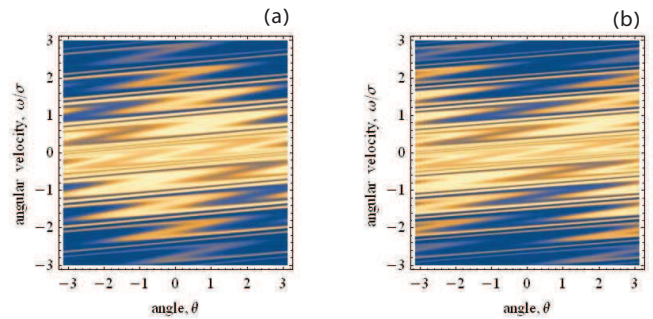}
\protect\caption{(color online) Echo formation in the filamented phase space. $\Omega_1 /\sigma =1$, $\sigma T=10$,  and $\Omega_2/\Omega_1=1/3$. Panel (a): $\sigma \tau = 9.11$ - orientation echo. Panel (b): $\sigma\tau = 10.85$ - anti-orientation echo.}
\label{joint_echo}
\end{figure}

Assume now that at  $t=T$, the ensemble of rotors is subject to the second impulsive stimulus.   The central filament in Fig.~\ref{firstkick}c (or Fig.~\ref{firstkick}d) which passes through the origin $(\theta ,\omega )= (0,0)$   is quite analogous  to the initial uniform distribution of rotors shown in Fig.~\ref{orientation} by the blue line, and it reacts similarly to the kick. With time, this filament forms the typical folded pattern  leading to the transient orientation of the rotors in the filament.   Other filaments experience the same transformation, with an important addition - for every angle $\theta$, the emerging folded patterns start moving with respect to the central filament with the velocity difference $l*2\pi /T$, where  $l$ is the filament number counted from the central one. As a result, the kick-induced patterns are generally shifted with respect to each other most  of the time after the kick, which results in a more or less uniform total angular distribution considered as a function of $\theta$ only.
If however the ensemble is observed with the delay $\tau \sim T$ after the kicking pulse, the kick-induced patterns of different filaments pile up on the top of each other in  phase space (see Fig.~\ref{joint_echo}a) manifesting  increased  density  near $\theta=0$   and an echo in the orientation factor $\langle \cos(\theta )\rangle$. Figure~\ref{joint_echo}b demonstrates  orientation along $\theta = \pi$ that happens a little bit later in the same time region (anti-orientation echo).  This time evolution of the filaments reveals the mechanism  leading to echoes in an ensemble of free rotors.  The process involves two stages: (i) the first short kick compresses a sizable fraction of the ensemble in the vicinity of $\theta=0$ (orientation/alignment effect), and the subsequent free  evolution  results  in filamentation  of the  phase  space density,  producing a series  of parallel strips separated in angular velocity by the integer multiples of $2\pi /t$ .  (ii) The second impulsive stimulus applied at a delay $T$ after the first one induces the orientation/alignment dynamics in every filament, and all the emerging orientation/alignment patterns synchronously overlap  near $\theta=0$  after another delay $\tau \sim T$  due to the above ``quantization" of the angular velocities of the strips.

\begin{figure}
\centering \includegraphics[clip=true,width=0.5\textwidth]{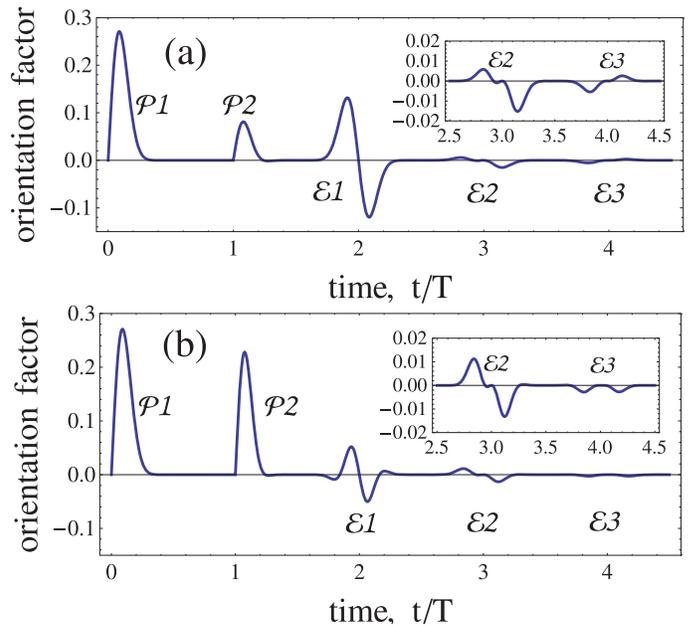}
\protect\caption{(color online) Orientation factor versus time after the first kick. $\Omega_1 /\sigma =1$ , $\sigma T=10$.   Panel (a): $\Omega_2/\Omega_1=1/3$. Panel (b): $\Omega_2/\Omega_1=1$. Here $\mathcal{P}1$ and $\mathcal{P}2$  denote  transient responses to the first and the second kick, respectively. $\mathcal{E}1$, $\mathcal{E}2$ and $\mathcal{E}3$ are echoes of the first, second, and third order, respectively. Insets show a magnified view of the second and third echoes. }
\label{echographs}
\end{figure}
These qualitative arguments have a considerable predictive power. In particular, it is clear that the same mechanism should form the echo signals also at delays $2T,3T, ...$  after the second pulse (higher order echoes). Moreover, it is expected that synchronization of the orientation/alignment patterns from non-neighboring strips at  $\tau = T/2, T/3$, ...  causes highly symmetric structures in the phase space, which may be associated with ``fractional echoes". These echoes are not seen in a mere orientation signal $\langle \cos(\theta )\rangle$, but require higher order observables $\langle \cos (n\theta )\rangle$ ($n > 1$) to be measured as a function of time. Moreover, just looking at the phase space pattern at the moment of the main echo, Fig.~\ref{joint_echo}a, one may expect that the echo is best manifested  if the delay $T$ coincides with the time needed for achieving the maximal orientation in every strip. This means that for a given delay there should be an optimal intensity of the second kick, so that the echo signal is maximal at this intensity value.

The simple 2D model  considered here (see Eq.(\ref{map})) allows for obtaining a fully analytical expression for the time-dependent mean value of $\langle \cos (n \theta )\rangle$ where $n$ is an integer:
\begin{multline}
\langle \cos (n \theta )\rangle (\tau )  = \\ \sum_{k=0}^{k=\infty}(-1)^k e^{-\frac{1}{2}\sigma ^2(n\tau -kT)^2} J_{k+n}[ \Omega_2 n\tau ]  J_k [\Omega_1 (n\tau -kT)]
  \label{echoanal}
\end{multline}
For $n=1,2$ this quantity is the orientation factor  and the alignment factor of the ensemble, respectively. Equation~(\ref{echoanal}) presents a sequence of signals localized in time near $\tau = \frac{k}{n} T$ where $k$ is an integer. For $n>1$, these are the above mentioned ``fractional echoes", while $n=1$ corresponds to the regular orientation echoes.
Figure~\ref{echographs} presents  calculated time-dependencies of the orientation factor for two values of the intensity of the second kick. It is clearly seen that the initial transient orientation shortly after the second pulse is followed by a series of echoes at $\tau \sim T, 2T$ and $3T$. In Figure~\ref{echographs}a, the intensity of the second kick, $\Omega_2$ was optimized to achieve the maximal amplitude of the main echo at $\tau \sim T$. Although  the second kick in Fig.~\ref{echographs}a is three times weaker than that of Fig.~\ref{echographs}b, it induces a considerably stronger echo, that even exceeds the  initial response to the second kick. The peak and dip of the first echo in Fig.~\ref{echographs}a correspond to the left and right panel of Fig.~\ref{joint_echo}, respectively. As follows from Eq.~(\ref{echoanal}), the amplitude of the echo is a decaying oscillatory function of the intensity of the second kick, $\Omega_2$ after reaching  the global maximum shown at Fig.~\ref{echographs}a. This is related to the oscillatory behavior of the Bessel function $J_{2}(\Omega_2 \tau )$ in Eq.~(\ref{echoanal}).

The same qualitative mechanism works in the case of the polarization-induced interaction (proportional to $-\cos^2(\theta )$) of the rotors with an external pulsed field, and it produces echoes in the ensemble averaged alignment signal, $\langle\cos^2(\theta ) \rangle$. An analytical expression for the alignment echo signals for 2D rotors is provided in the Supplementary Material~\cite{SM}, and it is quite similar to the Eq.(\ref{echoanal}).  The visualization of the phase space transformations in the case of alignment is more involved due to the doubling of the angular modulation of the filaments, however the main qualitative results remain the same, including the non-monotonous dependence of the echo signals on the intensity of the second kick.

In what follows, we describe the first experimental observation of the rotational alignment echoes in laser-kicked  CO$_2$ molecules.
The experimental setup had been used before \cite{Karras} and is described  in the Supplementary Material~\cite{SM}. The two linearly polarized pump pulses $\mathcal{P}1$ and $\mathcal{P}2$ are derived from an amplified Ti:Sapphire laser
(1 kHz, 800 nm, 100 fs FWHM) and properly delayed before being focused inside a CO$_2$ filled gas cell at a pressure of  0.2 bar.  The alignment echoes are observed by time-resolved birefringence measurements \cite{Renard03}, where a delayed, weak probe pulse, linearly polarized at +45$^\circ$  relative to the pump pulses is linearly analyzed after passing through the cell by a polarizer set at -45$^\circ$ relative to the pump pulses. The birefringence signal measured on the detector can be written as \cite{Faucher2011}
\begin{eqnarray}\label{expal}
\mathcal{S}(\tau)\propto\int_{-\infty}^{\infty} I_{\textrm{pr}}\left(t-\tau\right)\left[\langle \cos^2 (\theta)\rangle \left(t,I_1,I_2\right)-\frac{1}{3}\right]^2dt,
\end{eqnarray}
where $\langle \cos^2 (\theta)\rangle \left(t,I_1,I_2\right)$ is the  alignment factor,
and $I_1$, $I_2$, and $I_{\textrm{pr}}$,  are the intensity of  $\mathcal{P}1$,  $\mathcal{P}2$, and the probe pulse, respectively.

\begin{figure}[tb!]{}
  \begin{center}
    \includegraphics[width=8.3cm]{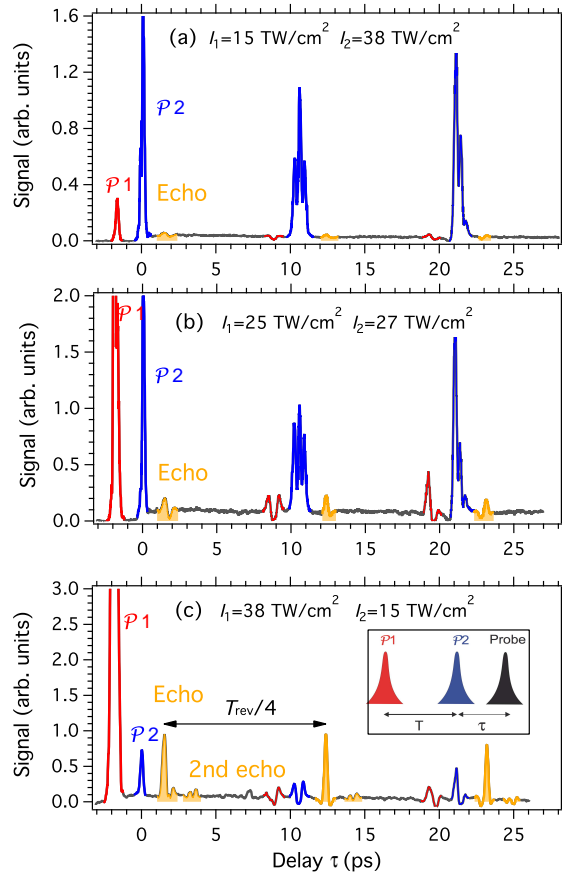}
  \end{center}
  \caption{(color online) Birefringence signals as a function of the pump-probe delay $\tau$ for different intensities $I_1$ and $I_2$  of the pump pulses $\mathcal{P}1$ and $\mathcal{P}2$, respectively (see the inset). $T_\textrm{rev}$: rotational revival time. }
  \label{IntDep}
\end{figure}

Figure \ref{IntDep} presents the pump-probe signals obtained for three different sets of pump intensities with a delay  $T$ between  $\mathcal{P}1$ and  $\mathcal{P}2$  set to 1.6 ps. We measured the alignment signals till about one-half of the rotational revival time, $T_\textrm{rev}$ of the CO$_2$ molecule ($T_\textrm{rev} \sim 42.7$ ps).  Classical echo effects discussed in this paper happen within the time interval of about $T_\textrm{rev}/4$ where the alignment dynamics is well described by the classical mechanics. To guide the eye, the alignment signals caused by the pulses $\mathcal{P}1$ and $\mathcal{P}2$, and also their one-quarter and one-half  revivals are colored in red and blue, respectively, whereas the echoes are highlighted  with a yellow color.
\begin{figure}[tb!]{}
  \begin{center}
    \includegraphics[width=8.3cm]{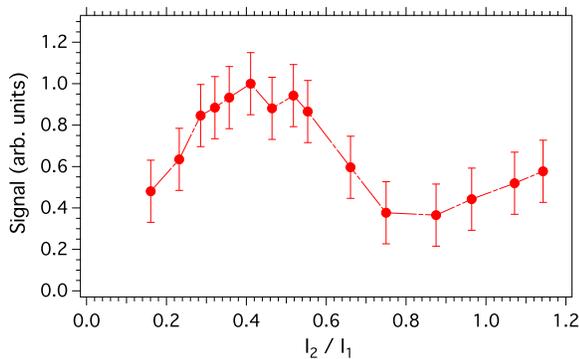}
  \end{center}
  \caption{(color online) Peak amplitude of the birefringence signal measured around the echo appearance near $\tau=T$ normalized to  its maximum value versus intensity ratio of the  pump pulses $I_2/I_1$,  with $I_1$ set to 44 TW/cm$^2$.}
  \label{EchoAmp}
\end{figure}
For each intensity set, an echo is produced  after $\mathcal{P}2$ at $\tau \sim T$.  Its amplitude depends on the intensities of both $\mathcal{P}1$ and $\mathcal{P}2$. The  maximum echo amplitude seen in Fig. \ref{IntDep}c is achieved for the  second pulse weaker than the first one,  $I_2/I_1\sim 0.4$. In this case the echo is stronger than the initial response induced by the second pulse $\mathcal{P}2$. These observations resemble the theoretical results shown in Fig. 4 considering the orientation process, and they may be understood using the similar qualitative arguments.  The second echo produced at  $\tau \sim 2T$ is also apparent in Fig. \ref{IntDep}c. On the longer time scale, replicas of these echoes are observed near the one-quarter and half-revival regions. They appear due to the interplay between the classical echo effect and the phenomenon of fractional quantum revivals \cite{fr}, and they are out of the scope of the present paper. These replicas are nicely reproduced in the fully quantum theoretical description of our experiments (to be published), and they have much in common with the revival echoes predicted in a collection of anharmonically confined cold atoms \cite{Fishman}. Related transient signals were reported in Ref.\cite{chinese} that studied  alignment structures of  N$_2$O molecules excited by two strong femtosecond laser pulses. However the delay between the pulses in these experiments was chosen close to the quarter-revival time, far beyond the region where the classical echo effects of the present paper exist.

We also performed intensity-dependent measurements  in order to reveal  the non-monotonic behavior of the echo amplitude with respect to the intensity of the second pump pulse $\mathcal{P}2$ while keeping the first one ($\mathcal{P}1$) fixed. The  results  are summarized  in Fig. \ref{EchoAmp} that clearly demonstrates the existence of an optimal intensity of the second kick, and  the oscillatory dependence of the echo amplitude on this intensity. Both observations are consistent with the presented theoretical considerations.

Summarizing, we showed   that a textbook classical system - a collection of randomly oriented free rotors with dispersed rotational velocities  - exhibits multiple echoes of different kinds in its orientation/alignment dynamics when kicked by a pair of delayed pulses. We provided qualitative and quantitative analysis of the phenomenon using a simple model system, and performed the first  experimental demonstration of the classical alignment echoes  in a thermal gas of CO$_2$ molecules excited by a pair of femtosecond laser pulses. The echoes were observed on a time scale much shorter than the quantum revival time, and their parameters were readily controllable by the intensity and delay of the excitation pulses. This makes the orientation/alignment echoes promising for exploring relaxation processes in high pressure  gases and various liquids, including superfluid helium~\cite{Helium}, where quantum rotational responses are considerably suppressed. Studies on  environmental influence on the orientation/alignment echoes are currently underway.

\acknowledgments
This work was supported by the Conseil R\'egional de Bourgogne (PARI program), the CNRS, the Labex ACTION program (contract ANR-11-LABX-01-01), and the French National Research Agency (ANR) through the CoConicS program (Contract No. ANR-13-BS08-0013).   This research was also supported by DFG (Project No. LE 2138/2-1) and Minerva Foundation. I.A. acknowledges kind hospitality and support from the Universit\'{e} Paris Est Cr\'{e}teil during a one week stay at LISA.

\widetext
\clearpage
 \renewcommand{\theequation}{S-\arabic{equation}}
  \setcounter{equation}{0}  
\begin{center}
      {\bf Supplementary material}
    \end{center}
\section*{Alignment echoes}
Here we briefly summarize results for alignment echo in the systems  that interact with  electromagnetic field via the induced polarization (\ref{polarizable}). The kick transformation describing the  angular velocity, $\omega$ and angle $\theta$ of a rotor at time $t$ after the first kick is given by the map (\ref{map}), in which $\sin (\theta_0 )$ is replaced by $\sin (2\theta_0 )$. By applying this transformation twice, we obtain the angular position of the rotor at time $\tau$ after the second kick that was applied at $t=T$:
\begin{equation} \label{angle}
\theta = \theta_0+\omega_0 T - \Omega_1 T \sin (2\theta_0) +\tau \left\{\omega_0  - \Omega_1  \sin (2\theta_0)-\Omega_2  \sin \left[2(\theta_0+\omega_0 T - \Omega_1 T \sin (2\theta_0))\right]\right\}
\end{equation}
Using this expression, we calculate the time-dependent alignment factor
\begin{equation} \label{alfactor}
\langle \cos^2 (\theta )\rangle= \frac{1}{2}+\frac{1}{2}\Re \langle e^{2i\theta}\rangle=\frac{1}{2}+\frac{1}{2(2\pi)^{3/2}}\Re\left[\int_{-\pi}^{\pi}d\theta_0 \int_{-\infty}^{+\infty}d\omega_0 e^{-\omega_0^2/(2\sigma^2)} e^{2i\theta}\right]
\end{equation}
With the help of the well known formula
\begin{equation}\label{Bessel}
e^{iz\sin (\theta )}=\sum_{k=-\infty}^{k=+\infty}e^{i k \theta}J_{k}(z),
\end{equation}
we expand $\exp\{-2i\tau \Omega_2  \sin \left[2(\theta_0+\omega_0 T - \Omega_1 T \sin (2\theta_0))\right]\}$ in series of Bessel functions $J_k(2\Omega_2 \tau )$, combine together all the terms proportional to $\omega_0$ in the exponent, and perform the Gaussian integration. This yelds
\begin{equation}\label{step1}
\langle \cos^2 (\theta )\rangle=\frac{1}{2}+\frac{1}{4\pi}\Re\sum_{k=-\infty}^{k=+\infty} J_k(2\Omega_2 \tau )e^{-2\sigma^2[\tau-(k-1)T]^2}\int_{-\pi}^{\pi}d\theta_0 e^{-2i\Omega_1 [\tau - (k-1)T]\sin (2\theta_0)-2i(k-1)\theta_0}
\end{equation}
We perform the remaining integration in (\ref{step1}) with the help of the inverse Fourier transform of Eq. (\ref{Bessel}), shift the summation index by 1, and neglect terms with the negative index (they are exponentially small if $\sigma T >1$):
\begin{equation}\label{final}
\langle \cos^2 (\theta )\rangle= \frac{1}{2}+\frac{1}{2}\sum^{k=\infty}_{k=0}(-1)^k e^{-2\sigma^2[\tau-k T]^2} J_{k+1}[2\Omega_2 \tau ]J_k [2\Omega_1 (\tau-k T)]
\end{equation}
Expression (\ref{final}) describes alignment echoes happening at $\tau\sim kT$ ($k=1,2,3 ...$). The term for $k=0$ describes the transient response just after the second pulse. The alignment factor (\ref{final}) is very similar (up to the overall shift by 1/2, and rescaling the parameters) to the analytical expression for the orientation factor (\ref{echoanal}). Therefore, all the qualitative properties of the echo signals discussed before for the orientation process, are applicable to the alignment process as well.
\section*{Experimental setup}

\begin{figure}[tb!]
  \begin{center}
    \includegraphics[width=8.5cm,keepaspectratio]{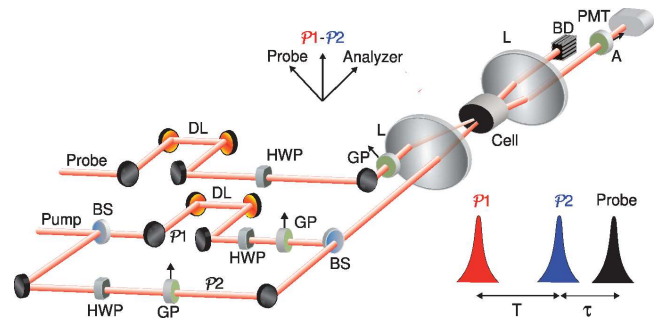}
  \end{center}
    \caption{(color online) Experimental set-up. A: Analyzer, BS: Beam Splitter, GP: Glan Polarizer, HWP: half wave-plate, PMT: Photo Multiplier Tube, L: Lens, DL: Delay Line, BD: Beam Dump. The relative polarizations of the different pulses  together with the orientation of the analyzer along with a relative timing chart are shown in  the insets.  Here $T$ is the delay between  $\mathcal{P}1$ and   $\mathcal{P}2$, and $\tau$ is the delay between $\mathcal{P}2$ and the probe pulse. }
  \label{exp_setup}
\end{figure}

A schematic representation of the experimental apparatus is presented in Fig. \ref{exp_setup}. The laser source is a Ti:Sapphire based, chirped-pulse amplified system delivering pulses centered at 800 nm with duration of 100 fs. The output pulses have a maximum energy of 3.5 mJ at 1 kHz repetition rate. 
A Mach-Zehnder interferometer generates  two pump pulses $\mathcal{P}1$ and $\mathcal{P}2$ with  a relative delay $T$ adjustable through a computer controlled translation stage equipped with a corner cube reflector. These two pulses are linearly polarized in the same direction and focused inside a static cell using a plano-convex lens. The cell is filled with CO$_2$ molecules at room temperature under a  pressure of 0.2 bar. The alignment echoes are  observed by  time-resolved birefringence measurements, a non-invasive method so far successfully applied to the detection of field-free molecular alignment \cite{Renard03}. A probe pulse is produced by inserting a beam splitter in the beam path just after the exit of the laser source. It is delayed using a second  motorized translational stage,  its   energy is about two orders of magnitude lower than those of the pump pulses and its   polarization is linear and set to +45$^\circ$ with respect to the polarization of the pump pulses. The transient  molecular alignment induced by the pump pulses is revealed by the depolarization of the  probe field recorded as a function of the delay $\tau$ between $\mathcal{P}2$ and the probe  pulse.  The depolarized field component is selected by  a linear analyzer located after the cell and set at $-45^\circ$ with respect to the  polarization of the pump pulses. The birefringence signal measured on the detector can be written as shown in Eq.(\ref{expal}) (see \cite{Faucher2011}). Finally, control over the energy of all beams is achieved using the combination of zero-order half wave-plates and Glan polarizers.

\end{document}